\documentstyle[12pt]{article}

\newcommand{\be}{\begin{equation}}
\newcommand{\ee}{\end{equation}}

\newcommand{\beba}{\begin{equation}\begin{array}{lcl}}
\newcommand{\eaee}{\end{array}\end{equation}}
\newcommand{\bea}{\begin{eqnarray}}
\newcommand{\eea}{\end{eqnarray}}
\newcommand{\ba}{\begin{array}}
\newcommand{\ea}{\end{array}}

\newcommand{\nnu}{\nonumber}

\newcommand{\td}{\tilde}

\newcommand{\norsl}{\normalsize\sl}
\newcommand{\ns}{\normalsize}
\newcommand{\norsc}{\normalsize\sc}
\def\a{\alpha}
\def\b{\beta}
\def\g{\gamma}
\def\c{\chi}
\def\d{\delta}
\def\e{\epsilon}
\def\ep{\varepsilon}

\def\m{\mu}
\def\n{\nu}
\def\o{\omega}
\def\p{\pi}

\def\th{\theta}

\def\r{\rho}
\def\s{\sigma}

\def\x{\xi}

\def\G{\Gamma}

\def\L{\Lambda}

\def\ch{{\cal H}}
\def\cl{{\cal L}}
\def\cf{{\cal F}}
\def\tr{{\rm tr}}
\begin{document}

\textheight 22cm
\voffset  -1cm
\begin{titlepage}

\vskip 4cm

\title{\vskip 1.4cm \Large\bf {Wormhole Effects on Yang-Mills Theory }}
\author{
\norsc A. Lukas \\
\\
\norsl  Physik Department\\
\norsl  Technische  Universit\"{a}t M\"{u}nchen\\
\norsl  D-85747 Garching, Germany\\     }

\date{}
\maketitle

\begin{abstract}
In this paper wormhole effects on $SO(3)$ YM theory are examined. The
wormhole wave functions for the scalar, the vector and the tensor expansion
modes are computed assuming a small gauge coupling which leads to an
effective decoupling of gravity and YM theory. These results are used to
determine the wormhole vertices and the corresponding effective operators
for the lowest expansion mode of each type. For the lowest scalar mode
we find a renormalization of the gauge coupling from the two point
function and the operators $\tr (F^3)$, $\tr (F^2\td{F})$ from the
three point function. The two point function for the lowest vector mode
contributes to the gauge coupling renormalization only whereas the lowest
tensor mode can also generate higher derivative terms.
\end{abstract}

\begin{picture}(5,2)(-300,-470)
\put(2.3,0){TUM--TH--165/94}
\put(2.3,-20){hep-th/9407037}
\put(2.3,-40){July 1994}
\end{picture}

\thispagestyle{empty}
\end{titlepage}
\setcounter{page}{1}
\section{Introduction}

Great interest has been devoted to the question of topology change in space
time and its effects on low energy field theories. In the framework of
Euklidean path integral for gravity wormholes appear as stationary points
which can cause a small baby universe to branch off from a large smooth
universe. In the semiclassical approximation the effect of these wormholes
on the low energy theory can be determined~\cite{haw1} resulting in an
infinite number of local operators, one for each type of wormhole. These
effective operators are multiplied by unknown parameters $\a$ which
describe the ``wormhole content'' of the universe and are unobservable
from the large universe point of view. It has been argued that this can
introduce a new uncertainty into physics~\cite{haw1,gi_st} in addition to
that of quantum mechanics.

However, using the Hartle--Hawking wave function of the universe~\cite{hh}
Coleman~\cite{col} has proposed a method to fix this uncertainty.
In his scenario a probability distribution for the $\a$'s is computed in an
effective theory consisting of the original theory and the wormhole induced
operators. He showed that the probability distribution is peaked on a
subspace of $\a$'s which corresponds to a vanishing cosmological constant.
Furthermore, restricting to this subspace one might be able to fix the $\a$'s
completely and hence all other constants of nature. Such a reasoning can
e.~g.~be used to explain why $\th_{QCD}$ takes a CP conserving value~\cite{CP}.
Though Coleman's argumentation has been criticized by several
authors~\cite{co_cr} it still appears to be very attractive because of its
ability to solve long outstanding theoretical problems.

Clearly, its application requires an understanding of the wormhole induced
operators and their relation to the wormhole states. Such an analysis has
been carried out for a conformally coupled scalar field~\cite{haw1}, for
fermions~\cite{lyons}, for gravitons~\cite{dow_la} and for
electromagnetism~\cite{dowker}. In this paper we are interested in
wormhole effects on YM theory which we study by looking at the simplest
gauge group $SO(3)$. We will not touch the questions related to the validity
of the semiclassical approximation. Instead we assume a wormhole scale
safely below the Planck scale. It should be stressed that the semiclassical
approximation for YM theory can differ substantially from electromagnetism.
This arises because for YM theory on a space with spatial topology $S^3$
gauge freedom allows for a nontrivial symmetric state around which the
expansion is carried out. E.~g.~in contrast to electrodynamics YM
perturbations do not decouple from gravity and in the $SO(3)$ case harmonics
up to spin 2 (tensor harmonics) appear in the expansion. However, here we
will assume a small gauge coupling which causes an effective decoupling
of gravity and YM theory and makes the equations analytically solvable.
In this approximation the general wormhole wave function solution of
the Wheeler--De Witt equation is determined for the scalar the vector and
the tensor perturbations. It is used to compute the wormhole vertex of the
lowest mode for each type of perturbation.

Symmetric Einstein-YM (EYM) theory on a space with spatial topology $S^3$
has been constructed in ref.~\cite{bmpv}. In ref.~\cite{bert_mour} this
construction has been applied to determine the Hartle--Hawking wave
function for the gauge group $SO(3)$. A generalization which includes
perturbations has been given in ref.~\cite{kklm}. In the next section
we will briefly review some results of these papers which we will need.
Our main goal is to find the wormhole vertex
\bea
 &&<0|A_{\m_1}(y_1)\cdots A_{\m_k}(y_k)|\Psi>\; = \nnu \\
 &&\quad\int d[h] d[A_0]\Psi[h,A_0]\int d[g]d[A]A_{\m_1}(y_1)\cdots
  A_{\m_k}(y_k)\exp (-S_E [g,A]) \; .
 \label{wv}
\eea
The second path integral on the RHS of this expression has to be carried
out over all 4--geometries (metric $g$) which are bounded by a three
sphere $S^3$ and approximate flat space at infinity. Analogously, the gauge
field $A$ should match its value $A_0$ on $S^3$ and tend to zero
asymptotically.
The exponent $S_E$ is the Euklidean action of these 4--configurations
which relate the baby universe state specified by the wave functional $\Psi$
to flat space. This wave functional will be determined in section 3.
In section 4 we describe the semiclassical approximation for the path
integral. Wormhole effects for the two- and three-point function induced
by the symmetric mode are discussed in section 5 and section 6 will
extend this to the lowest vector and tensor mode. We conclude in section 7.
Two appendices are also added which fix the geometry of $S^3$ as
a coset space and the conventions for harmonic expansion on it.

\section{Harmonic expansion of $SO(3)$ YM-theory}

In this section we present the basic formulae for the harmonic expansion
of $SO(3)$ YM--theory on a space with spatial geometry $S^3$ which have
been derived in ref.~\cite{kklm}. Our conventions concerning the geometry
of $S^3$ and the harmonics are summarized in appendix A.

The gauge field $A \equiv A_0 dt + A_a \omega^a$ is splitted into a
symmetric part $A^{(0)}$ and perturbations $\bar{A}$
\bea
 A &=& A^{(0)} + \bar{A}\\
 A^{(0)}_a &=& [1+\c(t)]T_a \; .
\eea
Here $A^{(0)}$ has the property that its change under
$G=SU_L(2)\times SU_R(2)$, the symmetry group of $S^3$, can be compensated
by a suitable gauge transformation. The expansion for $\bar{A}$ reads
\bea
 \bar{A}_0 &=& [\sqrt{2}\a_n S_b^n+\b_n P_b^n] T_b \label{exp0}\\
 \bar{A}_a &=& {1\over\sqrt{6}}\gamma_n \delta_{ab} Q^n T_b +
         \left[{1\over\sqrt{2}}\rho_n S_c^n
         +{1\over\sqrt{6}}\sigma_n P_c^n\right ] \ep_{acb} T_b \nnu \\
         &&+\mu_n G_{ab}^n T_b
         +{1\over\sqrt{2}}\nu_n S_{ab}^n T_b
         +\sqrt{6}\xi_n P_{ab}^n T_b \; . \label{exp}
\eea
The Lifshitz-harmonics are denoted by $Q,P,S$ and $G$ and we have chosen
a simplified notation where all relevant indices including even/odd are
hidden in $n$.
All expansion coefficients are constant on the coset and depend only on time.
They split into scalar--coefficients ($\g ,\x , \s, \b$), vector--coefficients
($\n , \r , \a$) and a tensor--coefficient ($\m$). These three types
completely separate from each other when the Lagrangian is computed up to
second order in the perturbations. Here we will not reproduce this
Lagrangian which has been determined in ref.~\cite{kklm} since in view of the
wormhole wave function we are mainly interested in the Hamiltonian.
To proceed in an explicit gauge invariant way we study the effect of
an infinitesimal gauge transformation
$\d A_\m = \nabla_\m U-[U,A_\m]$, $U=u_aT_a$ on the perturbations~:
\be
 \label{trafo}
 \ba{llllllllllll}
 \d\g_n&=&-\frac{1}{3}v_n \ ,
 &\d\x_n&=&\frac{1}{6}v_n \ ,
 &\d\s_n&=&\c v_n \ ,
 &\d\b_n&=&\dot{v}_n , \\
 \d\n_n&=&w_n \ ,
 &\d\a_n&=&\dot{w}_n \ \ ,
 &\d\r_n&=&2\c w_n+n\tilde{w}_n \ ,
 &\d\m_n&=&0 \, ,
\ea
\ee
where we have expanded
\be
 u_a = \frac{1}{\sqrt{6}}v_n P_a^n+\sqrt{2}w_nS_a^n\; .
 \label{u_exp}
\ee
These transformations suggest the introduction of gauge invariant variables
which are particular useful to formulate explicitly invariant wave functions.
A possible choice is
\be
 \label{inv}
 \ba{lllllllll}
 \Gamma_n&=&\g_n+2\x_n \ ,
 &B_n&=&\b_n+3\dot{\g}_n \ ,
 &S_n&=&\s_n+3\c\g_n \ , \\
 A_n&=&\a_n-\dot{\n}_n \,
 &R_n&=&\r_n-2\c\n_n-n\tilde{\n}_n \; . &&&
 \ea
\ee
For the $n=2$ scalar-- and vector--coefficients a special treatment
is necessary since neither $\x_2$ nor $\n_2$ appears in the Lagrangian.
We define
\be
 s = \frac{1}{e}\left( \frac{1}{3}\s_2 +\c\g_2\right)\; ,\quad\quad
 r = \frac{1}{e}(\r_2^{(e)}-\c\r_2^{(o)}) \; ,
 \label{inv2}
\ee
where $(e)$/$(o)$ refer to the even/odd component of a perturbation.

For the gravitational part we assume a metric of the form
\be
 g = \s^2(-N_0^2 dt^2 + a(t)^2 \sum_b \o^b \o^b)\; ,
 \quad\s^2=\frac{2}{3\p m_P^2}\; ,
 \label{metric}
\ee
i.~e.~we will neglect gravitational perturbations. Therefore, in addition
to the decoupling of the symmetric quantities $a$ and $\c$ by means
of conformal invariance of YM theory~\cite{bmpv} we demand a decoupling
of the corresponding perturbations. According to ref.~\cite{kklm}
this arises automatically if the field $\c$ is assumed to be close
to one of its ``classical'' values $\c = \pm 1$. This assumption is
well justified for a small gauge coupling $e$ in which case the wave
function for $\c$ is strongly peaked around these values~\cite{bm}
(see eq.~(\ref{wf_back}) below). Consequently, as long as perturbations are
concerned we assume that $\c=\bar{\c}=\pm 1$.

\section{Hamiltonian and Wormhole Wave Function}

Working with the unperturbed metric~(\ref{metric}) the Hamiltonian shows the
following structure~:
\be
 H = N_0(\ch_{0,Gr}+\ch_{0,YM}+\sum_n \ch_2^{(n)})
     +\sum_n(\a_n\ch_\a^{(n)}+\b_n\ch_\b^{(n)})\; .
 \label{ham_str}
\ee
In the gauge $N_0=a$ the Hamiltonians for the symmetric degrees of
freedom $a$ and $\c$ read
\beba
 \ch_{0,Gr} &=& \frac{\ns 1}{\ns 2}(-\p_a^2-a^2) \\
 \ch_{0,YM} &=& \frac{\ns e^2}{\ns 2}\left(\frac{\ns 1}{\ns 3}\p_\c^2+6V\right)
 \label{ham_back}
\eaee
with
\be
 V = \frac{1}{2e^4}(\c^2-1)^2 \; .
 \label{c_pot}
\ee
Using the linear Hamiltonian constraints enforced by the Lagrange
multipliers $\a_n$ and $\b_n$
\beba
 \ch_\a^{(n)}&=&2\bar{\c}\p_{\r_n}+\p_{\n_n}+n\tilde{\p}_{\r_n} \\
 \ch_\b^{(n)}&=&\bar{\c}\p_{\s_n}-\frac{\ns 1}{\ns 3}\p_{\g_n}
               +\frac{\ns 1}{\ns 6}\p_{\x_n}
 \label{ham_gau}
\eaee
it can be shown that the wave function should
only depend on the gauge invariant combinations defined in~(\ref{inv}) and
(\ref{inv2}). This fact can be used to rewrite the quadratic Hamiltonians
$\ch_2^{(n)}=^S\ch^{(n)}+^V\ch^{(n)}+^T\ch^{(n)}$ exclusively in terms of these
combinations~:
\bea
 ^T\ch^{(n)} &=& \frac{e^2}{2}\p_{\m_n}^2
           +\frac{1}{2e^2}[(n^2+1)\m_n^2+2n\bar{\c}\m_n\td{\m}_n] \nnu \\
 ^V\ch^{(n)} &=& \frac{2e^2 n^2}{n^2-4}\left[\frac{1}{2}\p_{R_n}^2
                 +\frac{\bar{\c}}{n}\p_{R_n}\td{\p}_{R_n}\right]+
                 \frac{1}{4e^2}[n^2 R_n^2-2n\bar{\c} R_n \td{R}_n] \label{ham}
\\
 ^S\ch^{(n)} &=& e^2\left[\frac{3(n^2-3)}{2(n^2-4)}\p_{\G_n}^2+
                 \frac{3}{2}(n^2+5)\p_{S_n}^2+6\bar{\c}
                 \p_{\G_n}\p_{S_n}\right] \nnu \\
              && +\frac{1}{6e^2}[(n^2-4)\G_n^2 + S^2] \; . \nnu
\eea
In these expressions a sum over even/odd is implied and the tilde operation is
defined by $\tilde{p}_n^{(even/odd)}=p_n^{(odd/even)}$ for a generic
perturbation $p_n$. The $n=2$ vector and scalar part has to be treated
separatly. In terms of the variables defined in~(\ref{inv2}) the Hamiltonians
take the simple form
\be
 ^V\ch^{(2)} = \p_r^2+r^2\; ,\quad\quad
 ^S\ch^{(2)} = \frac{3}{2}(\p_s^2+s^2) \; .
 \label{l_ham}
\ee
{}\\

We are now ready to compute the wormhole wave function $\Psi$ subject to
the Wheeler--De Witt equation
\be
 H\Psi = 0
 \label{wdw}
\ee
which enters the expression~(\ref{wv}) for the wormhole vertex.
It can be written as a product
$\Psi = \Psi_0 (a,\c) \Psi_2 (p)$ with wave functions $\Psi_0$ and
$\Psi_2$ for the symmetric fields and the perturbations, respectively.
Let us first concentrate on $\Psi_0$.
The only nonlinear term in $H$ is contained in the potential~(\ref{c_pot})
for the symmetric field $\c$. However, in the case of a small gauge
coupling which we assume here we can neglect tunneling between the two
minima at $\bar{\c}=\pm 1$. Then the field $\c$ sits close to one of
these minima and approximately feels the linearized potential
\be
 V_\pm = \frac{2}{e^4}(\c\mp 1)^2 \; .
\ee
Under this assumption we easily find a complete set of harmonic oscillator
solutions for~(\ref{ham_back})
\bea
 \Psi_{0,\pm k_{Gr}k_{S0}} &\sim& H_{k_{Gr}}(a) H_{k_{S0}}(\s_\pm)\exp
                   \left( -\frac{1}{2}(a^2+\s_\pm^2)\right) \label{wf_back}\\
 \s_\pm &=& \frac{\sqrt{6}}{e}(\pm\c -1) \; ,
\eea
where $H_k$ are the Hermitian polynomials and we have omitted the normalization
constant. The label $k_0$ is interpreted as the number of $\c$ quanta in
the wormhole. Their eigenvalues are given by
\be
 E_0 = -k_{Gr}+2k_{S0}+\frac{1}{2}\; .
 \label{eig_symm}
\ee

Now we turn to the perturbations. To simplify the notation from now on we
will stick to a certain mode and drop the index $n$. We write the separation
ansatz
\be
 \Psi_2 = ^S\Psi (\G,S)\; ^V\Psi (R_e,R_o)\; ^T\Psi (\m_e,\m_o)
 \label{wf_symm}
\ee
which automatically fulfills the two constraints~(\ref{ham_gau}). The
diagonalization of the kinetic terms and the potentials in the quadratic
Hamiltonians~(\ref{ham}) is straightforward. For $n>2$ we find the
eigenmodes
\bea
 \m_\pm &=& \frac{1}{e}\sqrt{\frac{n\pm\bar{\c}}{2}}(\m_e\pm\m_o) \nnu \\
 R_\pm &=& \frac{1}{2e}\sqrt{\frac{n^2-4}{n\pm 2\bar{\c}}}(R_e\pm R_o) \\
 \G_\pm &=& \frac{1}{e}\sqrt{\frac{n\pm 2}{6n(n\pm 1)}}
            ((n\mp 2)\G\pm\bar{\c} S) \nnu
\eea
corresponding to eigenvalues
\bea
 E_T &=& (k_{T+}+\frac{1}{2})(n+\bar{\c})
         +(k_{T-}+\frac{1}{2})(n-\bar{\c}) \nnu \\
 E_V &=& (k_{V+}+k_{V-}+1)n \label{eig_pert} \\
 E_S &=& (k_{S+}+\frac{1}{2})(n+1)+(k_{S-}+\frac{1}{2})(n-1) \nnu \; .
\eea
The wave function is given by harmonic oscillator solutions in the above
eigenmodes. The Wheeler-De Witt equation~(\ref{wdw}) is satisfied if all
energies sum up to zero~:
\be
 k_{Gr} = 2k_{S0}+k_{T+}(n+\bar{\c})+k_{T-}(n-\bar{\c})+(k_{V+}+k_{V-})n
     +k_{S+}(n+1)+k_{S-}(n-1) \; .
\ee
We have assumed that the ground state energy is canceled by a suitable
renormalization.

Again the case $n=2$ needs a special treatment~: The coefficients $r$ and $s$
in eq.~(\ref{inv2}) are already chosen as eigenmodes of the
Hamiltonian~(\ref{l_ham}) with energies
\be
 E_S = 3(k_S+\frac{1}{2})\; ,\quad\quad E_V = 2(k_V+\frac{1}{2})\; .
\ee

In the following we will be interested in wormholes which contain quanta
from the lowest mode in each the
scalar-, the vector- and the tensor sector. For the scalar sector this lowest
mode ($n=1$) is just the symmetric field $\c$. The lowest vector- and tensor
modes are the ones for $n=2$ and $n=3$, respectively. For later reference
we collect some of the above results to write the properly normalized
``lowest $n$'' wave function~:
\bea
 \Psi^{(Gr)}(a) &=& N(k_{Gr})H_{k_{Gr}}(a)\exp\left(-\frac{1}{2}a^2\right)
 \label{l_grav} \\
 \Psi^{(YM)}_{\pm k_S k_V k_+ k_-}(\s_\pm,r,\m_+,\m_-) &=&
      N H_{k_S}(\s_\pm)H_{k_V}(r)H_{k_+}(\m_+)H_{k_-}(\m_-) \nnu \\
      &&\times\exp\left(-\frac{1}{2}(\s_\pm^2+r^2+\m_+^2+\m_-^2)\right)
 \label{l_YM} \\
 N &=& N(k_S)N(k_V)N(k_+)N(k_-) \nnu
\eea
with the normalization constant
\be
 N(k) = \frac{1}{\p^{1/4}}\frac{1}{\sqrt{2^k k!}} \;
\ee
and $k_{Gr}$ given by
\be
 k_{Gr} = 2k_S+2k_V+(3+\bar{\c})k_+ +(3-\bar{\c})k_-\; .
\ee

\section{Semiclassical Approximation}

As a next step in the calculation of the wormhole vertex we have to
determine the saddle point of the path integral in eq.~(\ref{wv}).
For this the Euklidean equations of motion subject to appropriate
boundary conditions have to be solved. At conformal time $\eta=0$ the
solutions should match the geometry and the particle content specified
by the wormhole wave function whereas at $\eta\rightarrow\infty$
flat empty space should be approximated. For the scale factor $a$ and
the lowest $n$ YM modes these boundary conditions and the equations of
motion are summarized in table~\ref{boundary}.
\begin{table}
 \begin{center}
 \begin{tabular}{|c|c|c||c|}
  \hline
  & $\eta=0$ & $\eta\rightarrow\infty$ & EOM \\ \hline
  $a$ & $a_0$ & $\infty$ & $a'' = a$ \\ \hline
  $\c$ & $\c_0$ & $\pm 1$ & $\c ''=dV/d\c$ \\ \hline
  $r$ & $r_0$ & $0$ & $r'' =4r$ \\ \hline
  $\m_\pm$ & $\m_{0\pm}$ & $0$ & $\m_\pm '' =(3\pm\bar{\c})^2\m_\pm$ \\ \hline
 \end{tabular}
 \caption{Boundary conditions and equations of motion for the lowest modes}
 \label{boundary}
 \end{center}
\end{table}
The solutions are
\be
 a=a_0\exp (\eta)\; ,\quad\c_\pm =\pm 1\pm\frac{e}{\sqrt{6}}\s_{0\pm}
 \frac{a_0^2}{a^2}\; ,\quad r=r_0\frac{a_0^2}{a^2}\; ,\quad
 \m_\pm=\m_{0\pm}\left(\frac{a_0}{a}\right)^{(3\pm\bar{\c})}
\ee
leading to an Euklidean action of
\be
 S_{E,\pm} = \frac{1}{2}(a_0^2+\s_{0\pm}^2+r^2+\m_+^2+\m_-^2)\; .
 \label{eukl_act}
\ee
These solutions can be inserted into the harmonic expansion for the gauge
field~(\ref{exp0}) and (\ref{exp}). For the vector modes we have a gauge
freedom from the infinitesimal transformation~(\ref{trafo}). It allows us to
set $A_0=0$ and to write $A_a$ in the specific form
\bea
 A_{\pm ,a}(y) &=& \left[1\pm 1\pm\frac{e}{\sqrt{6}}\s_{0\pm}\frac{a_0^2}{a^2}
                   \right] T_a \nnu \\
                  &&+\frac{1}{4}er_{0q}\frac{a_0^2}{a^2}\left[ (1-\bar{\c})
                  D_{cq}^{(1|10)}(\s (y))+(1+\bar{\c})
                  D_{cq}^{(1|01)}(\s (y))\right]
                  \e_{acb}T_b \nnu \\
               && +e\left[\frac{\m_{0+,(pq)}}{\sqrt{3+\bar{\c}}}
                  \left(\frac{a_0}{a}\right)^{(3+\bar{\c})}
                  D_{(ab)(pq)}^{(2|20)}(\s (y))\right. \nnu \\
               &&\quad\quad \left. +\frac{\m_{0-,(pq)}}{\sqrt{3-\bar{\c}}}
                  \left(\frac{a_0}{a}\right)^{(3-\bar{\c})}
                  D_{(ab)(pq)}^{(2|02)}(\s (y))\right] T_b \label{a_exp} \; .
\eea
Here we have used the formulae~(\ref{l_S}) and (\ref{l_G}) from appendix A.
The sphere $S^3$ is thought of being embedded in 4--dimensional flat space
centered around $x_0$ and is parameterized by the four vector
$y$~: $a^2=(x_0-y)^2$. The correspondence between $y$ and a group element
$g\in SU(2)\cong S^3$ is fixed in appendix B.

Before we specify eq.~(\ref{wv}) for the wormhole vertex we remark that
the path integral has to be carried out over all insertion points $x_0$.
This integration ensures momentum conservation of the result. Furthermore we
note that all gauge fields which arise from eq.~(\ref{a_exp}) by a
$SU_L(2)\times SU_R(2)$ rotation are solutions of the equation of motion
with the same Euklidean action~(\ref{eukl_act}). Therefore an integration
over all these configurations has to be performed.
In order to remove the asymmetry between $A_{+,a}$ and $A_{-,a}$ which
differ by the pure gauge background $2T_a$ we carry out the gauge
transformation $A_+\rightarrow UA_+ U^{-1} -dUU^{-1}$ with $U=D^1(y)$ ($D^1$ is
the spin 1 representation of $SU(2)$). In addition this transformation
causes a change in the matrices $T_a$~: $T_a\rightarrow D^1_{ab}(y)T_b$.
Now we can write
\bea
 &&<0|\tr (A_{\m_1}(y_1)\cdots A_{\m_k}(y_k)|\pm k_Sk_Vk_+k_->\; = \nnu \\
 &&\quad \int da_0d\s_{0\pm}dr_0d\m_{0+}d\m_{0-}\Psi^{(Gr)}(a_0)
    \Psi^{(YM)}_{\pm k_Sk_Vk_+k_-}(\s_{0\pm} ,r_0,\m_{0+},\m_{0-}) \nnu \\
  &&\quad\times\int dx_0d\m (g_L)d\m (g_R)\prod_{i=1,k}P_{a_i\m_i}(x_0,y_i)
    D^1_{a_ib_i}(g_L^{-1}) \nnu \\
   &&\quad\times\tr (A_{\pm ,b_1}(gx_1)\cdots A_{\pm ,b_k}(gx_k))
    \exp(-S_{E,\pm})\; , \label{ver_1}
\eea
with $g=(g_L,g_R)$ and the wormhole wave function as given
in~(\ref{l_grav}), (\ref{l_YM}). The matrices $P_{a\m}$ rotate from a
coordinate system specified by the one forms $\o^a$ on $S^3$
to Euklidean coordinates in the flat 4-dimensional embedding space and
are specified in appendix B. Rotating by $D^1(g_L^{-1})$ is necessary
because of the transformation property~(\ref{w_trafo}) of $\o^a$.

In the following we will evaluate this expression for the various modes.
This allows us to determine the effective interactions
$\cl_{\pm k_Sk_Vk_+k_-}$ in the asymptotic flat space which satisfy
\bea
 &&<0|\tr (A_{\m_1}(y_1)\cdots A_{\m_k}(y_k))|\pm k_Sk_Vk_+k_->\; \sim \nnu \\
 &&\quad\int d^4 x_0 <0|\tr (A_{\m_1}(y_1)\cdots A_{\m_k}(y_k))
    \cl_{\pm k_Sk_Vk_+k_-}(x_0)|0>\; . \label{eff_int}
\eea
The gauge choice in eq.~(\ref{a_exp}) guarantees that the path integral
is performed over fields with
$\partial_\m A_\m^b = \partial_\m P_{a\m}A_a^b = 0$. The flat space amplitude
on the RHS of eq.~(\ref{eff_int}) should therefore be evaluated in a
covariant gauge with a gauge fixing term $(\partial_\m A_\m^a)^2$. Clearly,
the requirement $\partial_\m A_\m^b = 0$ does not eliminate the whole gauge
freedom. For simplicity this remaining freedom has been chosen such that
$A_0$ vanishes.

\section{Wormhole Vertices from the Symmetric Mode $\c$}
An evaluation of eq.~(\ref{ver_1}) for a wormhole state containing only
quanta in the symmetric mode $\c$ leads to
\bea
 &&<0|\tr (A_{\m_1}(y_1)\cdots A_{\m_k}(y_k)|\pm k_S=k>\; = \nnu \\
  &&\quad\quad (\pm)^k C(k)\int dx_0\; \tr (T_{a_1}\cdots T_{a_k})
               \prod_{i=1,k}\frac{ P^\pm_{a_i\m_i}(p_i)}{p_i^2}
  \label{ver_s}
\eea
with $p_i = x_0-y_i$ and the numerical constants $C(k)$ which result from the
first integration in eq.~(\ref{ver_1}) over the wormhole degrees of
freedom. The matrices $P^\pm$ are defined in appendix B. We can concentrate
on the case with equal number of quanta and insertion points ($k_S=k$).
If $k_S >k$ the wormhole vertex vanishes by means of integration over
the wormhole wave function. All quanta of the wormhole state have to be
created or annihilated in the asymptotically flat region~\cite{haw1}. In
the opposite case contractions in flat space appear which we are not
interested in.

To calculate the integrand of eq.~(\ref{ver_s}) for the 2- and 3-point function
let us define the states
\be
 |e/o,k_S> = \frac{1}{\sqrt{2}}(|+k_S>\pm |-k_S>) \; .
\ee
They are even $(e)$ and odd $(o)$ with respect to the parity $\hat{P}_2$
on $S^3$. Asymptotically this parity approaches the ordinary parity in
flat space. With the help of eq.~(\ref{p_2}) and (\ref{p_3}) we get
\bea
 <0|\tr (A_\m (x)A_\n (y))|e/o, k_S=2> &=& \frac{e^2V_{S^3}^2}{\sqrt{2}}C(2)
    \left\{ \begin{array}{cc}
            I_2(x,y) & {\rm (e)} \\
            0 & {\rm (o)}
    \end{array} \right. \\
 I_2(x,y) = \int dx_0\frac{1}{p^2q^2}(p.q\d_{\m\n}-p_\n q_\m )&&
\eea
and
\bea
 &&<0|\tr (A_\m (x)A_\n (y)A_\n (z))|e/o, k_S=3>\; = \nnu \\
 &&\quad\quad\quad\quad\quad
   \frac{e^3V_{S^3}^2}{2}C(3)\int dx_0\frac{1}{p^2q^2s^2}\times
         \left\{ \begin{array}{cc}
         (p.sq_\m -p.qs_\m) & {\rm (e)} \\
         (\e_{\m\n\r\s}p_\n q_\r s_\s) & {\rm (o)}
         \end{array} \right. \; .
\eea
The effective interactions which reproduce these amplitudes are
\bea
 \cl_{e,k_S=2} &\sim& \tr (F_{\m\n}F_{\m\n}) \nnu \\
 \cl_{e,k_S=3} &\sim& \tr (F_{\m\n}F_{\n\r}F_{\r\m}) \\
 \cl_{o,k_S=3} &\sim& \tr (F_{\m\n}F_{\n\r}\td{F}_{\r\m}) \; . \nnu
\eea
The parity even lowest scalar mode causes a renormalization of the gauge
coupling $e$. It shows the same behaviour as the lowest modes of
electrodynamics~\cite{dowker}. From the 3-point function we find the two
higher operators which can be written in terms of the field strength only.
Their behaviour under parity reflects the parity of the corresponding
wormhole wave function.

\section{Effects of the Lowest Vector-- and Tensor Mode}
Finally, we are interested in the 2--point functions which result from the
lowest vector mode $r_0$ and the lowest tensor modes $\m_{0+}$, $\m_{0-}$.

The integration over $SU_L(2)\times SU_R(2)$ in eq.~(\ref{ver_1}) causes
a contraction of the non--coset indices of the harmonics and the additional
indices of the perturbations, e.~g.~leading to $\sum_{q=1,2,3}r_{0q}r_{0q}$.
Therefore, if the two quanta needed for the 2--point function are excited
in different components of $r_{0q}$ (or $\m_{0+,q}$, $\m_{0-,q}$) the
wormhole vertex vanishes by means of orthogonality of the Hermitian
polynomials. Taking both quanta in the same component we obtain for
the vector mode
\be
 <0|\tr (A_\m (x)A_\n (y))|e/o, k_V=2> = \frac{e^2}{\sqrt{2}}C(2)
    \left\{ \begin{array}{cc}
            I_2(x,y) & {\rm (e)} \\
            0 & {\rm (o)}
    \end{array}\; , \right.
\ee
where we have defined eigenfunctions of the parity $\hat{P}_2$
\be
 |e/o,k_V> = \frac{1}{\sqrt{2}}(|+k_V>\pm |-k_V>)
\ee
as we did in the scalar case. To arrive at this result we have used the
explicit form of the harmonics~(\ref{l_harm}) and appendix B.
Again only the parity even modes contribute
to the renormalization of the gauge coupling. For a different (but covariant)
gauge choice as that fixed by eq.~(\ref{a_exp}) we could have generated
an additional contribution to the gauge fixing term $(\partial_\m A_\m^2)^2$,
i.~e.~a renormalization of the gauge fixing parameter. The simplicity of this
result is due to the fact that for $\bar{\c}=1$ ($\bar{\c}=-1$) only the pure
right handed (left handed) vector harmonics appear in the
expansion~(\ref{a_exp}). If deviations of $\bar{\c}$ from $\pm 1$ are taken
into account higher dimensional corrections to this result can be expected.

Even with out such generalizations the tensor modes show a less trivial
behaviour. The states $|\pm , k_\pm =2 , k_\mp = 0>$ will project the terms
with $(x_0-y)^{-4}$ out of the expansion~(\ref{a_exp}) and lead to higher
derivative operators. The other states
\be
 |e/o, k_T=2> = \frac{1}{\sqrt{2}}(|+k_-=2k_+=0>\pm |-k_+=2k_-=0>)\; ,
\ee
however, show the familiar behaviour
\be
 <0|\tr (A_\m (x)A_\n (y))|e/o, k_T=2> = \frac{5e^2}{3\sqrt{2}}C(2)
    \left\{ \begin{array}{cc}
            I_2(x,y) & {\rm (e)} \\
            0 & {\rm (o)}
    \end{array}\; . \right.
\ee
which can be seen by using~(\ref{l_harm}) and appendix B.

\section{Conclusion}

In this paper we have computed the wormhole wave function for $SO(3)$
YM theory in the case of a small gauge coupling. In that case we found
it to be well approximated by harmonic oscillator wave functions.

This wave function has been used to determine the wormhole vertex and the
corresponding flat space effective interaction for the lowest mode of each
type of perturbation. We found the parity behaviour of these effective
interactions to reflect the $S^3$ parity behaviour of the wave function.
In particular, only the parity even wave functions generate a renormalization
of the gauge coupling from the 2--point function. While this renormalization
exhausts the effects from the lowest scalar and vector 2--point function
the lowest tensor mode can also contribute to higher dimensional operators
with additional derivatives. In addition, we calculated the 3--point
function for the lowest scalar coefficient (the symmetric mode).  We
found the operators $\tr (F^3)$ and $\tr (F^2\td{F})$ from the parity even
and parity odd wave function, respectively.

The minisuperspace wave function of the universe for $SO(3)$ YM theory as it
has been proposed
in ref.~\cite{bert_mour} shows a degeneracy in the two minima $\c=\pm 1$
of the symmetric mode $\c$ which results from parity invariance. This
might give rise to an indeterminacy in fixing the constants of nature in
the sense of Coleman. However, this fixing has to be performed in an
effective theory taking into account all wormhole induced interactions.
One can hope that due to the parity violation of the second operator above
such a degeneracy can be lifted.

A departure from a small gauge coupling would complicate the situation
substantially since gravity and YM expansion modes would be coupled.
We have not addressed this situation here. Clearly, the wave function
would be much less simpler in such a case. Concerning the wormhole vertex
one can expect effective operators coupling gravity and YM with each other
which might be an interesting aspect for future studies.\\[1cm]

{\bf Acknowledgement}~This work was partially supported by the Deutsche
Forschungsgemeinschaft and the EC under contract no.~SC1-CT92-0789 and
the CEC Science Program no.~SC1-CT91-0729.
\renewcommand{\theequation}{\thesection.\arabic{equation}}
\newcommand{\sect}[1]{\section{#1}\setcounter{equation}{0}}
\section*{Appendices}
\appendix
\sect{Coset Geometry and Harmonic Expansion on $S^3$}

In this appendix we will briefly develop the geometry of $S^3$ as a coset
space and discuss harmonic expansion on this space. In particular the
relation between the various types of harmonics used in the
literature~\cite{camp,lif,dow_pett} will become transparent.

To follow the formalism of ref.~\cite{camp} we identify
$S^3$ with the coset $G/H$ where
$G=SU_L(2)\times SU_R(2)=\{g=(g_L,g_R)|g_L,g_R\in SU(2)\}$ and
$H=SU_D(2)=\{(h,h)| h\in SU(2)\}$. The representatives of the coset
elements are fixed by the map $\s$~: $\s((g_L,g_R)H) = (e,g_R g_L^{-1})$.
This definition immediately shows how to identify $S^3$ with $SU(2)$.
We will use this identification in the following to denote a coset
element represented by $(e,y)$ just as $y\in SU(2)$. An element
$g=(g_L,g_R)\in G$ acts on the coset by left multiplication which we write
as $\r_g$ or sometimes simply as $g$. A change of the representatives
induced by such a multiplication can be compensated by a function $\cf$
defined by $g\s(y)=\s(gy)\cf (g,h)$. For our particular choice of $\s$ we
find $\cf (g,y)=(g_L,g_L)$.

The Lie--Algebra of $G$ can be splitted into
the Lie--Algebra of $H$ spanned by $T_a=T_a^{(L)}+T_a^{(R)}$ and
an orthogonal coset part spanned by $T^{(c)}_a=T_a^{(L)}-T_a^{(R)}$.
Here $T^{(L)}$ and $T^{(R)}$ satisfy the usual commutation relations
$[T_a^{(L,R)},T_b^{(L,R)}]=\e_{abc}T_c^{(L,R)}$ and are normalized by
$\tr (T_a^{(L,R)}T_b^{(L,R)}) = 2\d_{ab}$.
An analogous decomposition can be applied to the Maurer--Cartan form of $G$
resulting in the one forms $L^a(g)$ and $L^{(c)a}(g)$ which are mapped
to one forms $e^a(y)=\s^* L^a(\s(y))$, $e^{(c)a}(y)=\s^* L^{(c)a} (\s(y))$
on the coset. Their properties crucially
depend on $\s$ and $\cf$. For the above choice we find that both types of
forms are equal $\o^a(y):=e^{(c)a}(y)=e^a(y)$ and satisfy the Maurer--Cartan
equation $d\o^a+{\e^a}_{bc}\o^b\wedge\o^c =0$. Under left multiplication $\o^a$
transforms as
\be
 \r_g^* \o^a(y) = D^1(g_L)^a_b\, \o^b(gy) \; ,
 \label{w_trafo}
\ee
where $D^j$ denotes the spin $j$ representation of $SU(2)$ and $g=(g_L,g_R)$.
Tensors on the coset will be given with respect to the forms $\o^a$.
Finally, the connection on $G/H$ is given by
$\o_{abc}=\e_{abc}$.\\

In ref.~\cite{kklm} the correspondence between Lifshitz--harmonics~\cite{lif}
and the harmonics derived from the representation matrices of $G$~\cite{camp}
has been established up to spin $J=2$. For the scalar harmonics and the
purely transverse vector and tensor harmonics it has been found that
\bea
Q_q^{(n)}(y) &=& {n^2} ~D^{(0|\frac{n-1}{2},\frac{n-1}{2})}_{0q}(\s(y)) \\
S^{(n,e/o)}_{a|q}(y) &=& \sqrt{{n^2-1} \over {6} }
\left[ D^{(1|\frac{n}{2},\frac{n}{2}-1)}_{aq}(\s(y)) \pm
       D^{(1|\frac{n}{2}-1,\frac{n}{2})}_{aq}(\s(y)) \right] \\
G_{a_1 a_2|q}^{(n,e/o)}(y) &=& \sqrt{ {n^2-4} \over {10 } }
\sum_{q_1,q_2}
P(a_1 a_2|q_1 q_2) \nnu \\
&&\times\left[ D^{(2|\frac{n+1}{2},\frac{n-3}{2})}_{(q_1 q_2)q}(\s(y))\pm
D^{(2|\frac{n-3}{2},\frac{n+1}{2})}_{(q_1 q_2)q}(\s(y)) \right]\; .
\eea
where the projection operator $P$
\be
P(a_1 a_2|q_1 q_2) \equiv [{1 \over 2} (\delta_{q_1 a_1} \delta_{q_2 a_2}
+ \delta_{q_2 a_1} \delta_{q_1 a_2})
-{1 \over 3} \delta_{q_1 q_2} \delta_{a_1 a_2}].
\ee
is needed to render the tensor harmonic $G$ symmetric and traceless in
its two coset indices.
The matrices $D_{mq}^{(J|j_L,j_R)}$ are $(j_L,j_R)$ representations of $G$
where the first index $m$ only runs over the part corresponding to the
$H$ representation $(J)\subset (j_L,j_R)$. The harmonics $S$ and $G$ split
into even $(e)$ and odd $(o)$ type with respect to the parity
$\hat{P}_2$~\cite{lif} which corresponds to the ordinary flat space parity. It
can be thought of as exchanging $SU_L(2)$ and $SU_R(2)$. The other Lifshitz
harmonics with longitudinal directions can be expressed as
\bea
P_{a|q}^{(n)}(y) &=& {1 \over {n^2-1}} \nabla_a Q^{(n)}_q (y) \\
P_{a_1 a_2|q}^{(n)}(y) &=& {1 \over {n^2-1}} \nabla_{a_1} \nabla_{a_2}
Q^{(n)}_q(y)
+{1 \over 3} \delta_{a_1 a_2} Q^{(n)}_q(y) \\
S_{a_1 a_2|q}^{(n,e/o)}(y) &=& {1 \over 2} [\nabla_{a_1} S_{a_2|q}^{(n,e/o)}(y)
+\nabla_{a_2} S_{a_1|q}^{(n,e/o)}(y)] \; .
\eea

To get explicit expressions for the matrices $D^{(J|j_L,j_R)}$ which
fit to our conventions for the coset geometry we start with a
$G$ representation
\be
 D^{(j_L,j_R)}_{(ns)(pq)} (g_L,g_R) = \left( D^{j_L}_{np}(g_L),
                                    D^{j_R}_{sq}(g_R) \right) \; .
\ee
The spins $(j_L,j_R)$ have to be coupled in the first pair $(ns)$ of
indices. In addition we use the definition of the representatives fixed
by the map $\s$ and obtain
\bea
 D^{(J|j_L,j_R)}_{m(pq)}(\s(y))&=&\sum_{n,s} <Jm|j_L n\, j_R s>
                             D^{j_L}_{np}(e),D^{j_R}_{sq}(y)\\
   &=&\sum_n <Jm|j_L p\, j_R n> D^{j_R}_{nq}(y) \; .
\eea
In this particular form they correspond to the harmonics of
ref.~\cite{dow_pett} which have been used by Dowker~\cite{dowker} to
compute wormhole effects on electrodynamics.

Their transformation property under $G$ can be determined from the above
definitions~:
\be
 D^{(J|j_L,j_R)}_{m(pq)}(\s (gy)) = D^J_{mm'}(g_L)D^{j_L}_{pp'}(g_L)
                           D^{j_R}_{qq'}(g_R)D^{(J|j_L,j_R)}_{m'(p'q')}
                           (\s (y))\; .
\ee
As a last step to make contact with Lifshitz harmonics one has to rotate
back from the Cartan basis for $SU(2)$ representations to a ``tensor'' basis.
We will do this explicitly for the lowest purely transverse vector and
tensor harmonics
\bea
 S_{a|q}^{(2,e/o)}(y) &=& \frac{1}{\sqrt{2}}(D_{aq}^{(1|10)}(\s (y))\pm
                          D_{aq}^{(1|01)}(\s (y))) \label{l_S} \\
 G_{ab|pq}^{(3,e/o)}(y) &=& \frac{1}{\sqrt{2}}(D_{(ab)(pq)}^{(2|20)}(\s (y))\pm
                          D_{(ab)(pq)}^{(2|02)}(\s (y)))
 \label{l_G}
\eea
which we will need in our calculation. We find
\beba
 D_{ap}^{(1|10)}(y) &=& \d_{ap} \\
 D_{ap}^{(1|01)}(y) &=& D^1_{ap}(y) \\
 D_{(ab)(pq)}^{(2|20)}(y) &=& P(ab|pq) \\
 D_{(ab)(pq)}^{(2|02)}(y) &=& \sum_{r,s} P(pq|rs)D^1_{ar}(y)D^1_{bs}(y)\; .
 \label{l_harm}
\eaee

\sect{Rotation Matrices and Covariant Expressions}
In the following we will establish the relation between the coordinates
of $S^3$ specified by the one forms $\o^a$ and the Cartesian coordinates
of the 4-dimensional embedding space.

A four vector $(p_\m)$ is mapped to a group element of $SU(2)\cong S^3$ by
\be
 (p_\m)\rightarrow \frac{1}{p^2}\left( \begin{array}{cc}
                                       p_4+ip_3 & p_2+ip_1 \\
                                       -p_2+ip_1 & p_4-ip_3
                                \end{array} \right) \; .
 \label{id}
\ee
The forms $\o^a$ can be calculated in a particular coordinate system
$(\x^\a)$ with $\o^a = \o^a_\a (\x)d\x^\a$. If $\L^\a_\m (p,\x)$ denotes the
transformation matrix from $(\x^\a)$ to Cartesian coordinates the desired
transformation relating vectors in the basis $\o^a$ to four vectors is given
by $P^a_\m (p) = \L^\a_\m (p,\x)\o_\a^a (\x)$. The calculation
has been carried out in ref.~\cite{dowker} and results in
\beba
 P_{ai} (p) &=& \frac{1}{p^2}(\e_{aji}p_j +\d_{ai}p_4 ) \\
 P_{a4} (p) &=& -\frac{1}{p^2}p_a \; .
\eaee
The spin 1 representations of $SU(2)$ can be expressed in terms of $(p_\m)$
using the identification~(\ref{id})~:
\be
 D^1_{ab}(p) = \frac{1}{p^2}((2p_4^2 -p^2)\d_{ab}+2p_a p_b
               + 2p_4\e_{abc}p_c ) \; .
\ee
This allows us to explicitly handle the harmonics of eq.~(\ref{l_harm}).
With the definitions $P^-(p) = P(p)$ and $P^+(p) = D^1(p)P(p)$ we obtain
\beba
 P^\pm_{ai}(p) &=& \frac{1}{p^2}(\mp\e_{aji} p_j +\d_{ai}p_4 ) \\
 P^\pm_{a4}(p) &=& -\frac{1}{p^2}p_a \; .
\eaee
Contracting the coset indices of the matrices $P^\pm$ in an $SU(2)$ invariant
manner should lead to covariant expressions. In particular
\bea
 P^\pm_{a\m}(p) P^\pm_{a\n}(q) &=& \frac{1}{p^2q^2}(p.q\d_{\m\n}-p_\n q_\m
                                   \mp\e_{\m\n\r\s}p_\r q_\s )
                                   \label{p_2} \\
 \e_{abc}P^\pm_{a\m}(p)P^\pm_{b\n}(q)P^\pm_{c\n}(s) &=&
                          \frac{2}{p^2q^2s^2}(\pm (p.sq_\m -p.qs_\m )
                          +\e_{\m\n\r\s}p_\n q_\r s_\s ) \; .
                          \label{p_3}
\eea

\end{document}